\documentclass[aps,prl,twocolumn,noshowpacs,showkeys]{revtex4}
\usepackage{graphicx,amsfonts,amssymb,amsmath,bm}

\begin{document}

\title{Low-temperature anomaly in heat capacity \\
due to overlapping the spectrums in molecular crystals}

\author{N.A.Nemov} \email{nine@che.nsk.su}
\author{V.N.Naumov}
\author{G.I.Frolova}
\author{V.R.Belosludov}
\author{V.N.Ikorskii}
\author{M.A.Bespyatov}
\affiliation{A.V.Nikolaev Institute of Inorganic Chemistry, Russian Academy of Sciences, Siberian Depart.,\\ Prospekt Academika Lavrentyeva 3, Novosibirsk 630090, Russia}

\begin{abstract}
Investigations of dynamic and thermodynamic properties for a molecular crystal tris-hexa\-fluoro\-acetylacetonate-iron $Fe(O_2C_5HF_6)_3$ are presented. Heat capacity $C_p(T)$ has been measured by adiabatic calorimetry method in the temperature range $4.8-321\ K$. An anomaly with a maximum at $T_c = 44.6\ K$ has been discovered. Intermolecular vibrations spectrum was calculated by lattice dynamics method in quasiharmonic approximation. Intramolecular frequencies are found by solving the Schr\"odinger equation in approach of small harmonic oscillations. In the frequency interval $\approx 30-70\ cm^{-1}$ overlapping the spectrums intra- and intermolecular oscillations has been found. The good agreement for calculated and experimental $C_p(T)$ occurred to be possible for two sets of the force constants. These sets describe two phases above and below $T_c$. The difference between phases is connected with freezing of rotation $CF_3$ groups. It has been concluded that the interaction between different modes leads to phase transition and anomaly in heat capacity. 
\end{abstract}

\date{\today}
\maketitle


\section{introduction}
Modern  materials technology even more often interferes with study of complicated molecular crystals containing hundreds atoms in a unit cell. A description of molecular crystals is usually based on an idea of intermolecular and intramolecular degrees of freedom. As a rule, in simple molecular crystals the spectrums of inter- and intramolecular vibrations are energy separated. In complicated molecular crystals these spectrums can overlap. Then a possible interactions between different types of vibrations cause an occurring of new phenomena (resonances, phase transitions and others). 

In this work the results of investigation of dynamic and thermodynamic properties for the molecular crystal tris-hexafluoroacetylacetonate-iron $Fe(O_2C_5HF_6)_3$ (or $Fe(hfac)_3$, $\beta$-diketonate $Fe$) are presented. 


\section{experiment}
\begin{figure}
\includegraphics[width=0.45\textwidth]{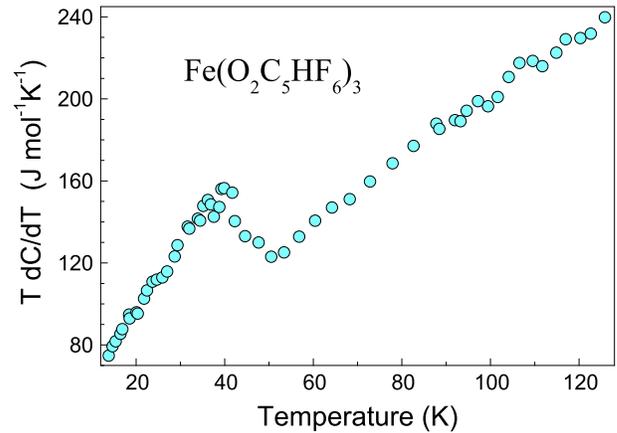}
\caption{\label{fig1} Experimental heat capacity of $Fe(hfac)_3$ in coordinates $T dC(T)/dT$ vs $T$ in the region of anomaly.}
\end{figure}

Heat capacity $C_p(T)$ of a sample $Fe(hfac)_3$ has been measured by adiabatic procedure in the temperature range $4.8 - 321\ K$. In the stepwise heating mode, 141 experimental points of the heat capacity have been obtained \cite{Naum04}. After examining the $C_p(T)$ dependence, the anomalous behavior of heat capacity has been discovered in the temperature range $30 - 60\ K$. It can be clearly seen on the plot of derivative $dC_p(T)/dT$, Fig.~\ref{fig1}. The amplitude of an extracted anomaly make up $\approx 3\%$ from regular heat capacity at $T_c = 44.6\ K$ (experimental error is equal $0.16\%$ within this temperature range).

\begin{figure}
\includegraphics[width=0.45\textwidth]{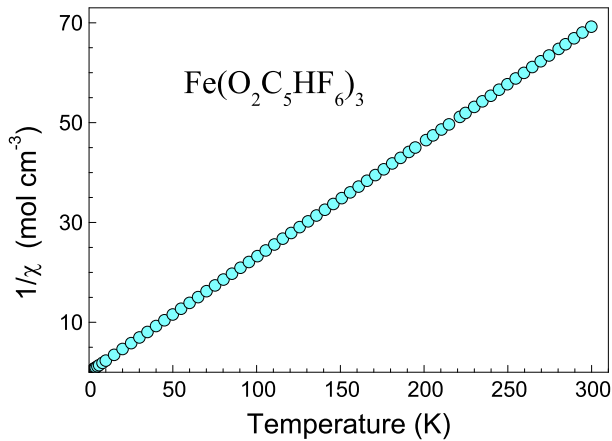}
\caption{\label{fig2} Inverse static magnetic susceptibility $1/\chi$ as a function of temperature.}
\end{figure}

To understand the nature of anomaly the static magnetic susceptibility was measured by MPMS-5s SQUID-magnetometer (Quantum Design) within the temperature range $2 - 300\ K$. The sample $Fe(hfac)_3$ has shown paramagnetic behavior, see Fig.~\ref{fig2}. Experimental points coincide well with the Curie-Weiss law, $\chi(T) = C_K/ \left( T - \Theta_K \right)$, where Curie constant $C_K = (4.331 \pm 0.003)\ cm^3K mol^{-1}$, and paramagnetic temperature $\Theta_K = (- 0.1951 \pm 0.0002)\ K$. No pronounced magnetic anomalies were observed. One can say, the anomaly in heat capacity has nonmagnetic nature.

\section{intermolecular vibrations}
Computer calculation of thermodynamic properties, connected with intermolecular oscillations of $Fe(hfac)_3$, was carried out within the framework of lattice dynamics method. Earlier this approach has been applied for description of dynamic and thermodynamic properties for $\beta$-diketonates of metals in the harmonic approximation \cite {BelHarm}. In this paper more complicated procedure based on quasiharmonic approach \cite{BelAnharm} has been used. A component part of this method is prestress optimization of structure. This is caused by lack of information about crystal structure and, in the greater degree, about interatomic potentials and their parameters. Detailed description of optimization procedure and extended possibilities of lattice dynamics method for molecular crystals had been published in Ref. \cite{OptBel2000}. 

\begin{figure}
\includegraphics[width=0.45\textwidth]{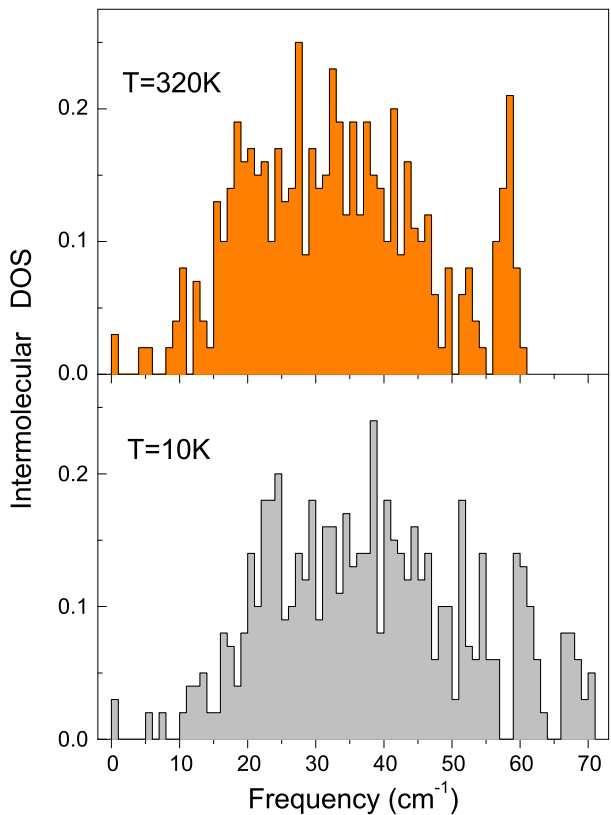}
\caption{\label{fig3} Intermolecular vibrational density of states at $320\ K$ and $10\ K$. The normalization of curves is made on 6.}
\end{figure}

The general plan of calculation looks as follows. Initial data are the experimental parameters of atom-atomic potentials, a unit cell and positions of molecules. At first the geometry optimization of crystal is carried out. As a result, all atoms occupy equilibrium positions at predetermined pressure. Then the dynamic matrix is calculated. Eigenvalues of this matrix determine oscillation frequencies of a crystal and allow one to calculate thermodynamic properties.

Free energy of the crystal $F_{qh}$ in that approach is calculated as $F_{qh} =U + F_{vib}$, where $U$ is the potential energy and $F_{vib}$ is vibrational part of free energy
    \[
    F_{vib}  = \frac{1}{2} \sum_{j,\vec q} \hbar \omega _j (\vec q) +  k T \sum_{j,\vec q} \ln \left(1 - e^{-{\hbar \omega _j (\vec q)}/{k T}} \right).
\]
Here $k$ is the Boltzmann constant, $\omega_j (\vec q)$ are the lattice vibrational frequencies. In quasiharmonic approximation the free energy for a given crystal structure is taken as for harmonic approximation. The anharmonisity causes mode frequency dependence of the structural parameters, making $F_{vib}$ a function of these parameters as well as of temperature. This dependence can be deduced by self-consistent manner as a result of optimizing a crystal structure while calculating a free energy.

In this work  molecules of a crystal $Fe(hfac)_3$ are considered as rigid. The unit cell given in Ref.\cite {struct-01} was used as initial structure with parameters $a = (9.057 \pm 0.004)$\AA, $b = (13.424 \pm 0.005)$\AA, $c = (21.591 \pm 0.016)$\AA, $\alpha = \gamma = 90^{\circ}$, $\beta = (116.71 \pm 0.02)^{\circ}$ for $T = 320\ K$. A unit cell contains four molecules $Fe(hfac)_3$. 

For describing the interactions between atoms of different molecules the Buckingham potential as earlier in Ref. \cite {BelHarm} was used. In Ref. \cite{UFF} the universal representation of interatomic interaction potentials, Universal Force Fields, has been suggested. In terms of Universal Force Fields description the Buckingham potential looks as
     \[
    U(x) = \frac{D}{\zeta-6}\left[6\; e^{\zeta(1-x)} - \frac{\zeta}{x^6}\right] ,
\]
where $x=r/\rho$, $r$ is the interatomic distance, $\rho$ is the interatomic distance at the minimum of the potential (i.e., where $U(1) = -D$), $D$ is the potential well depth and $\zeta$ is a steepness factor. As is conventionally done, the general $\rho_{mn}$, $D_{mn}$ and $\zeta_{mn}$ for different $m$ and $n$ types of atoms are obtained from homonuclear parameters through the use of combination rules $\rho_{mn}=(\rho_m \rho_n)^{1/2}$, $D_{mn}=(D_m D_n)^{1/2}$ and $\zeta_{mn}=(\zeta_m \zeta_n)^{1/2}$. 
The potential parameters used in present work were taken from Ref. \cite{param1} for light atoms and from Ref. \cite{param2} for iron. In the mentioned above representation they are given in the Table~\ref{tab1}.

\begin{table}
\caption{\label{tab1} Universal Force Fields parameters of Buckingham potential for $Fe(hfac)_3$.}
\begin{ruledtabular}
\begin{tabular}{cccc}
Atom type, $n$   &   $\rho_n$ (\AA)   &   $D_n\ (kJ\ mol^{-1}$)  &  $\zeta_n$ \\
\hline
H  & 2.794 & 0.127 & 11.987  \\
C  & 3.804 & 0.332 & 14.001  \\
O  & 3.253 & 0.512 & 13.598  \\
F  & 3.237 & 0.218 & 14.922  \\
Fe & 3.957 & 1.140 & 13.611  \\
\end{tabular}
\end{ruledtabular}
\end{table}

\begin{figure}
\includegraphics[width=0.45\textwidth]{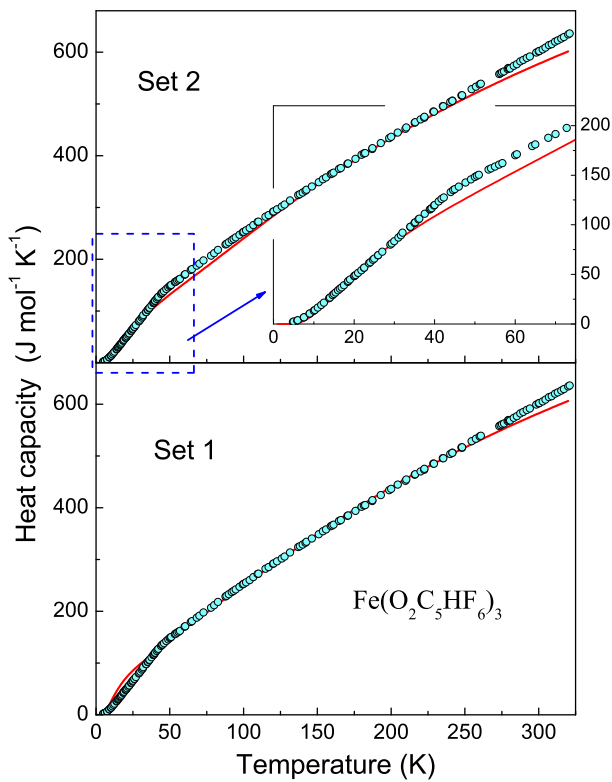}
\caption{\label{fig4} Heat capacity connected with intramolecular vibrations. Heat capacity without intermolecular contribution $C_{in}(T)$ is shown by circles. Straight-line is heat capacity calculated on bases of intramolecular vibrations for two Sets of force constants.}
\end{figure}

The shape of the cell and the positions of molecules were optimized for normal pressure, $P = 1\ bar$, at given temperature according to the technique described above. These parameters were used as initial ones to calculate structural parameters at lower temperatures. Further on, the procedure was repeated with stepwise temperature decreasing till $T = 1\ K$. Frequencies $\omega _j (\vec q)$ and density of phonon stats (DOS) have been calculated at every temperature using 5x5x5 points of the Brillouin zone. 

\begin{figure*}
\includegraphics[width=0.75\textwidth]{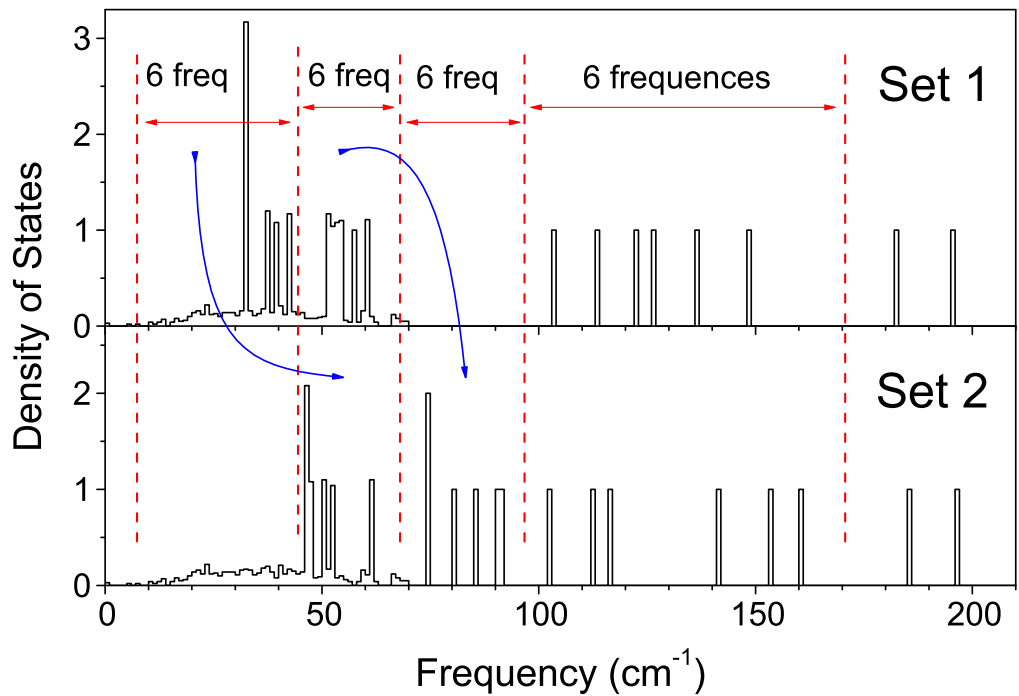}
\caption{\label{fig5} Low-frequency, $q < 210\ cm^{-1}$, total density of states for two Sets of force constants. Set 1 describes a phase for $T > T_c$, Set 2 corresponds to $T < T_c$ phase. Intermolecular DOS has 6 frequency modes and calculated for $45\ K$. Arrows show redistribution of intramolecular frequencies for Set 1 and Set 2.}
\end{figure*}

As a result of these calculations, intermolecular vibrational spectrums and relevant thermodynamic functions at different $T$ within the range $1 - 320\ K $ have been received. Intermolecular DOS for $10\ K$ and $320\ K$ is shown in Fig.~\ref{fig3}. The normalization of curves is made on 6, quantity intermolecular modes of one molecule. It is seen that as the temperature decreases from $320$ to $10\ K$ the edge of intermolecular vibration spectrum is uniformly shifted from $\approx 60$ to $\approx 70\ cm^{-1}$. Two localized bands separated by $6 - 8\ cm^{-1}$ can be observed near the edge of spectrum. As the temperature varies the marked bands are moved. They can consequently take on any values in the mentioned frequency range. 

On the base of intermolecular vibrational spectrum the corresponding component of heat capacity $C_{mol}(T)$ has been calculated within the range $1 - 320\ K$. These data allow to calculate $C_{in}(T)$, the contribution to a heat capacity depending from the intramolecular oscillations. For energy separated intra- and intermolecular vibrational
spectrums there is the relation $C_{in}(T) = C_p(T) - C_{mol}(T)$. Calculated in such a way $C_{in}(T)$ is shown in Fig.~\ref{fig4} by circles.


\section{intramolecular vibrations}
The vibrational spectrum of a molecule $Fe(hfac)_3$ is found in framework of classical description of small harmonic oscillations of atoms near their equilibrium positions in molecule. The complex of programs based on methods and algorithms published in Ref. \cite{Gribov3} has been used. The Schr\"odinger equation for a motion of 43 atoms with a potential energy $U_{pot} = \frac{1}{2} \sum_{i,j}u_{ij} q_i q_j $ has been solved. Here $u_{ij}$ are the force constants of interatomic interaction connected with vibrational coordinates $q_i$ and $q_j$. Initial parameters have been structure, characteristics of atom positions in the molecule \cite{struct-01} and force constants. Well-known force constants \cite{Yurch-eng, Ik-data-eng} that provide a good agreement of the calculated vibrational spectrum with experimental ones were used for valence bonds and angles. They were not varied further on. The constants $u_{ij}$ connected with non-planar coordinates and unknown for $Fe(hfac)_3$ were varied. The spectroscopic data on structural-isomeric compounds of this class as well as calculated \cite{31-eng} data for $\beta$ - diketonate chromium $Cr(O_2C_5H_7)_3$ were also used.

In standard method of choosing force constants (based on agreement of experimental and calculated spectroscopic data) complications arise when finding force constants responsible for the low-frequency spectral region $\leq 150 - 200\ cm^{-1}$. These problems are due to lack of any information about frequency spectrum in this region.  Instead of it, however, it is possible to use experimental data on heat capacity in the low-temperature interval $T \leq 320\ K $. In the present work was used new method for choosing force constants. The criterion for optimality has been the best agreement of \textit{functional behavior} of calculated on the bases of intramolecular vibrations heat capacity with $C_{in}(T)$. 

As a result of calculation, 123 frequencies responsible for interior oscillations of a molecule are found. It was discovered, there are two sets (Set 1 and Set 2) of force constants gives a good agreement calculated heat capacity and $C_{in}(T)$. First Set gives a good agreement of heat capacities in the temperature range $T > T_c$ (see Fig.~\ref{fig4}, Set 1). In the temperature range $T < T_c$ a good agreement takes place with the second Set (Fig.~\ref{fig4}, Set 2). A disagreement is comparable with experimental error. The systematic error above $250\ K$ may comes from anharmonic vibrations of intermolecular spectrum. 


\section{discussion}
Two finding various sets of force constants are appeared in different functional behavior of a heat capacity in areas above and below $T_c$. It gives the basis to suppose that these solutions may describe two different structural phases. More detailed difference between them can be seen on Fig.~\ref{fig5}. Here is shown the total DOS for two Sets of force constants in low-frequency region, $q < 210\ cm^{-1}$. One can see that the intramolecular spectrum is separated on parts containing 6 frequencies. It is marked by vertical dashed lines in Fig.~\ref{fig5}. For Set 1,  an overlapping intermolecular spectrum (6 modes) and intramolecular spectrum (12 modes) is realized. In $T < T_c$ phase, Set 2, intermolecular spectrum overlapped only with 6 intramolecular modes. The principle difference in spectrums takes place only in the region below $\approx 170\ cm^{-1}$. Just these modes effect essentially on the heat capacity and low temperature thermodynamic characteristics. The vibration modes over $\approx 600\ cm^{-1}$ contribute less then $\approx 1\%$. The contribution of vibration modes over $\approx 1200\ cm^{-1}$ is comparable with experimental error of $C_p(T)$. 

More exactly we can say that in phase $T < T_c$ (Set 2) in spectrums overlapping takes a part an intramolecular modes responding for low-frequency bending oscillations of $(O_2C_5HF_6)$-plane. Another 6 modes was displaced in more high-frequency area and do not take a part in overlapping of spectrums. These modes are formed by oscillations of $CF_3$ groups. These kind of oscillations may proceed to rotational type under increasing of amplitude of oscillations. In other words, it is possible that  at $T < T_c$ groups $CF_3$ finish to make rotational movements and there is their ordering.


\section{conclusion}
A new method based on using experimental heat capacity data for deciding of force constants allows us to take into account a correct density of modes on low frequency region of spectrum which would be problematic on the base of spectroscopic data. We find the overlapping intra- and intermolecular spectrums in complex molecular crystal. 
Spectrums are shifted due to temperature decreasing and it is possible that due to resonant interaction between spectrums arises two different structural phases above and below $T_c$. The difference between phases is connected with freezing of rotation $CF_3$ groups. And finding out anomaly in heat capacity is the consequence of spectrums overlapping.

A theoretical modeling and examining the experimental data obtained in this work and earlier \cite{31-eng} for $\beta$ - diketonate chromium lead us to conclusion that overlapping the spectrums of intra- and intermolecular vibrations is the characteristic property of molecular crystals $\beta$ - diketonate of metals. Certainly our consideration is not deprived the lacks occuring from originally separate consideration intra- and intermolecular fluctuations. More perfect technique of simultaneous description of coplex crystals is necessary.


\end{document}